\documentstyle[twoside]{article}
\catcode`\@=11
\long\def\@makefntext#1{
\protect\noindent \hbox to 3.2pt {\hskip-.9pt
$^{{\eightrm\@thefnmark}}$\hfil}#1\hfill}               

\def\@makefnmark{\hbox to 0pt{$^{\@thefnmark}$\hss}}    

\def\ps@myheadings{\let\@mkboth\@gobbletwo
\def\@oddhead{\hbox{}
\rightmark\hfil\eightrm\thepage}
\def\@oddfoot{}\def\@evenhead{\eightrm\thepage\hfil
\leftmark\hbox{}}\def\@evenfoot{}
\def\sectionmark##1{}\def\subsectionmark##1{}}

\oddsidemargin=\evensidemargin
\newcounter{sectionc}\newcounter{subsectionc}\newcounter{subsubsectionc}
\renewcommand{\section}[1] {\vspace{12pt}\addtocounter{sectionc}{1}
\setcounter{subsectionc}{0}\setcounter{subsubsectionc}{0}\noindent
        {\tenbf\thesectionc. #1}\par\vspace{5pt}}
\renewcommand{\subsection}[1] {\vspace{12pt}\addtocounter{subsectionc}{1}
        \setcounter{subsubsectionc}{0}\noindent
        {\bf\thesectionc.\thesubsectionc. {\kern1pt \bfit #1}}\par\vspace{5pt}}
\renewcommand{\subsubsection}[1] {\vspace{12pt}\addtocounter{subsubsectionc}{1}
        \noindent{\tenrm\thesectionc.\thesubsectionc.\thesubsubsectionc.
        {\kern1pt \tenit #1}}\par\vspace{5pt}}
\newcommand{\nonumsection}[1] {\vspace{12pt}\noindent{\tenbf #1}
        \par\vspace{5pt}}

\newcounter{appendixc}
\newcounter{subappendixc}[appendixc]
\newcounter{subsubappendixc}[subappendixc]
\renewcommand{\thesubappendixc}{\Alph{appendixc}.\arabic{subappendixc}}
\renewcommand{\thesubsubappendixc}
        {\Alph{appendixc}.\arabic{subappendixc}.\arabic{subsubappendixc}}

\renewcommand{\appendix}[1] {\vspace{12pt}
        \refstepcounter{appendixc}
        \setcounter{figure}{0}
        \setcounter{table}{0}
        \setcounter{lemma}{0}
        \setcounter{theorem}{0}
        \setcounter{corollary}{0}
        \setcounter{definition}{0}
        \setcounter{equation}{0}
        \renewcommand{\thefigure}{\Alph{appendixc}.\arabic{figure}}
        \renewcommand{\thetable}{\Alph{appendixc}.\arabic{table}}
        \renewcommand{\theappendixc}{\Alph{appendixc}}
        \renewcommand{\thelemma}{\Alph{appendixc}.\arabic{lemma}}
        \renewcommand{\thetheorem}{\Alph{appendixc}.\arabic{theorem}}
        \renewcommand{\thedefinition}{\Alph{appendixc}.\arabic{definition}}
        \renewcommand{\thecorollary}{\Alph{appendixc}.\arabic{corollary}}
        \renewcommand{\theequation}{\Alph{appendixc}.\arabic{equation}}
        \noindent{\tenbf Appendix \theappendixc #1}\par\vspace{5pt}}
\newcommand{\subappendix}[1] {\vspace{12pt}
        \refstepcounter{subappendixc}
        \noindent{\bf Appendix \thesubappendixc. {\kern1pt \bfit #1}}
        \par\vspace{5pt}}
\newcommand{\subsubappendix}[1] {\vspace{12pt}
        \refstepcounter{subsubappendixc}
        \noindent{\rm Appendix \thesubsubappendixc. {\kern1pt \tenit #1}}
        \par\vspace{5pt}}

\newcommand{\textlineskip}{\baselineskip=13pt}
\newcommand{\smalllineskip}{\baselineskip=10pt}

\def\eightcirc{
\begin{picture}(0,0)
\put(4.4,1.8){\circle{6.5}}
\end{picture}}
\def\eightcopyright{\eightcirc\kern2.7pt\hbox{\eightrm c}}

\newcommand{\copyrightheading}[1]
        {\vspace*{-2.5cm}\smalllineskip{\flushleft
        {\footnotesize International Journal of Modern Physics B, #1}\\
        {\footnotesize $\eightcopyright$\, World Scientific Publishing
         Company}\\
         }}

\newcommand{\publisher}[2]{{\begin{center}\footnotesize\smalllineskip
        Received #1\\
        Revised #2
        \end{center}
        }}

\def\abstracts#1#2#3{{
        \centering{\begin{minipage}{4.5in}\baselineskip=10pt\footnotesize
        \parindent=0pt #1\par
        \parindent=15pt #2\par
        \parindent=15pt #3
        \end{minipage}}\par}}

\renewenvironment{thebibliography}[1]                   
        {\frenchspacing
         \ninerm\baselineskip=11pt
         \begin{list}{\arabic{enumi}.}
        {\usecounter{enumi}\setlength{\parsep}{0pt}
         \setlength{\leftmargin 12.7pt}{\rightmargin 0pt} 
         \setlength{\itemsep}{0pt} \settowidth
        {\labelwidth}{#1.}\sloppy}}{\end{list}}

\newcounter{itemlistc}
\newcounter{romanlistc}
\newcounter{alphlistc}
\newcounter{arabiclistc}

\newenvironment{romanlist}
        {\setcounter{romanlistc}{0}
         \begin{list}{$($\roman{romanlistc}$)$}
        {\usecounter{romanlistc}
         \setlength{\parsep}{0pt}
         \setlength{\itemsep}{0pt}}}{\end{list}}

\newcommand{\fcaption}[1]{
        \refstepcounter{figure}
        \setbox\@tempboxa = \hbox{\footnotesize Fig.~\thefigure. #1}
        \ifdim \wd\@tempboxa > 5in
           {\begin{center}
        \parbox{5in}{\footnotesize\smalllineskip Fig.~\thefigure. #1}
            \end{center}}
        \else
             {\begin{center}
             {\footnotesize Fig.~\thefigure. #1}
              \end{center}}
        \fi}

\newcommand{\tcaption}[1]{
        \refstepcounter{table}
        \setbox\@tempboxa = \hbox{\footnotesize Table~\thetable. #1}
        \ifdim \wd\@tempboxa > 5in
           {\begin{center}
        \parbox{5in}{\footnotesize\smalllineskip Table~\thetable. #1}
            \end{center}}
        \else
             {\begin{center}
             {\footnotesize Table~\thetable. #1}
              \end{center}}
        \fi}

\def\@citex[#1]#2{\if@filesw\immediate\write\@auxout
        {\string\citation{#2}}\fi
\def\@citea{}\@cite{\@for\@citeb:=#2\do
        {\@citea\def\@citea{,}\@ifundefined
        {b@\@citeb}{{\bf ?}\@warning
        {Citation `\@citeb' on page \thepage \space undefined}}
        {\csname b@\@citeb\endcsname}}}{#1}}

\newif\if@cghi
\def\cite{\@cghitrue\@ifnextchar [{\@tempswatrue
        \@citex}{\@tempswafalse\@citex[]}}
\def\citelow{\@cghifalse\@ifnextchar [{\@tempswatrue
        \@citex}{\@tempswafalse\@citex[]}}
\def\@cite#1#2{{$\null^{#1}$\if@tempswa\typeout
        {IJCGA warning: optional citation argument
        ignored: `#2'} \fi}}

\def\pmb#1{\setbox0=\hbox{#1}
        \kern-.025em\copy0\kern-\wd0
        \kern.05em\copy0\kern-\wd0
        \kern-.025em\raise.0433em\box0}

\def\fnt#1#2{\footnotetext{\kern-.3em
        {$^{\mbox{\scriptsize #1}}$}{#2}}}

\def\fpage#1{\begingroup
\voffset=.3in
\thispagestyle{empty}\begin{table}[b]\centerline{\footnotesize #1}
        \end{table}\endgroup}

\def\runninghead#1#2{\pagestyle{myheadings}
\markboth{{\protect\footnotesize\it{\quad #1}}\hfill}
{\hfill{\protect\footnotesize\it{#2\quad}}}}
\font\tenrm=cmr10
\font\tenit=cmti10
\font\tenbf=cmbx10
\font\bfit=cmbxti10 at 10pt
\font\ninerm=cmr9
\font\nineit=cmti9
\font\ninebf=cmbx9
\font\eightrm=cmr8

\def\qed{\hbox{${\vcenter{\vbox{                        
   \hrule height 0.4pt\hbox{\vrule width 0.4pt height 6pt
   \kern5pt\vrule width 0.4pt}\hrule height 0.4pt}}}$}}

\def\bsc{{\sc a\kern-6.4pt\sc a\kern-6.4pt\sc a}}       
\def\bflatex{\bf L\kern-.30em\raise.3ex\hbox{\bsc}\kern-.14em
T\kern-.1667em\lower.7ex\hbox{E}\kern-.125em X}

\begin{document}

\runninghead{GENERALIZED SUPERSYMMETRIES}
{GENERALIZED SUPERSYMMETRIES}

\normalsize\textlineskip
\thispagestyle{empty}
\setcounter{page}{1}

\copyrightheading{}                     

\vspace*{0.88truein}

\fpage{1}
\centerline{\bf GENERALIZED SUPERSYMMETRIES
}
\vspace*{0.035truein}
\centerline{\bf AND COMPOSITE STRUCTURE  IN $M$-THEORY\footnote{Supported by KBN 
grant 5PO3B05620} }
\vspace*{0.37truein}
\centerline{\footnotesize JERZY LUKIERSKI}
\vspace*{0.015truein}
\centerline{\footnotesize\it Institute for Theoretical Physics, University
of Wroc{\l}aw }
\baselineskip=10pt
\centerline{\footnotesize\it pl. M. Borna 9, 50-205 Wroc{\l}aw, Poland
}
\vspace*{10pt}
\vspace*{0.225truein}
\publisher{(received date)}{(revised date)}

\vspace*{0.21truein}
\abstracts{ 
We describe generalized $D=11$
Poincar\'{e}
 and conformal supersymmetries. The corresponding
generalization of twistor and supertwistor framework is outlined
with $OSp(1|64)$ superspinors describing BPS preons.
 The $\frac{k}{32}$ BPS states as composed out of $n=32 - k$
 preons are introduced, and basic ideas  concerning BPS preon
  dynamics is presented. The lecture is based on results
obtained by J.A. de Azcarraga, I. Bandos, J.M. Izquierdo and the
author$^1$. 
}{}{}



\vspace*{1pt}\textlineskip      
\section{Introduction}    
\vspace*{-0.5pt} \noindent $M$-theory has been proposed as a
hypothetical quantum theory describing elementary level of
matter, which should incorporate and possibly explain various
properties of ``new string theory" (for review see e.g.$^{2-4}$%
). One of the features of such new theory of fundamental
interactions should be the appearance of many extended elementary
objects ($p$-(super)branes, $D$-(super)branes etc.) related with
each other via duality/dimensional reductions net. Such a variety
of basic objects in the theory makes sensible a search for some
underlying composite structure.

The basic dynamical degrees of freedom in $M$-theory yet are not
known - there were presented only some  proposals
 usually related with $D=11$ space-time geometry. We
postulate that the composite structure of $M$-theory should be
formulated in terms of  new degrees of freedom related
 with new geometry. Because
$M$-theory is supersymmetric, and supersymmetry reveals more
elementary nature of spinorial objects, we shall postulate that
the basic fundamental geometric structure in $M$-theory is
spinorial.

The only well-known part of the description of $M$-theory is
algebraic. Assuming that $M$-theory lives in $D=11$ ( this
assumption is consistent with description of $D=11$ SUGRA as the
low energy limit of $M$-theory) we can postulate the following
basic $D=11$ $M$-superalgebra\footnote{The relation (1) is the
standard, minimal $M$-superalgebra. One can also add arbitrary
spin-tensor central charges (see e.g.$^{5-6}$%
). The most general case was considered by Sezgin$^{7}$%
.}
\begin{equation}\label{lpr1.1}
  \left\{ Q_r, Q_s \right\} = Z_{rs} (\Gamma_\mu C )_{rs} P^k
  + C\Gamma_{[\mu\nu]} C)_{rs}
  Z^{[\mu\nu]} +
  \left( \Gamma_{[\mu_1 - \mu_s ]} C \right)_{rs}
  Z^{[\mu_1 \ldots \mu_5]}\, .
\end{equation}
where $\mu,\nu=0,1,\ldots 10,$ $r,s=1, \ldots 32$.
The collection of 528 Abelian generators $Z_{rs}$ ($Z_{rs}=
Z_{sr}$) describes the generalized momenta in $M$-theory.
Introducing dual generalized coordinate space
\begin{equation}\label{lpr1.2}
  X_{rs} = \left( \Gamma_{\mu} C\right)_{rs}
  X^{\mu} + \left( \Gamma_{[\mu\nu]} C\right)_{rs}
  \, X^{[\mu\nu]} +
  \left( \Gamma_{[\mu_5 - \mu_s]} C\right)_{rs} \, X^{[\mu_1
  \ldots \mu_5]}\, ,
\end{equation}
we obtain large generalized phase space, with coordinates and
positions described by the adjoint representations of $Sp(32)$
algebra.

Let us recall  the assumption of Penrose twistor formalism
 in $D=4$ $^{8-11}$  that basic spinorial degrees of
freedom in twistorial theory of elementary particles are described
by $N$ twistors $(i=1 \ldots N)$
\begin{equation}\label{lpr1.3}
 t^{(i)}_{\alpha} =
  \left( \lambda^{(i)}_{A}, \omega^{(i)\dot{A}}\right) \, ,
\end{equation}
where $\lambda_{A}^{(i)}$, 
$\omega^{(i)\dot{A}}$ ($A=1,2$) are the
pairs of $D=4$ Weyl spinors. The following formula for the
composite fourmomentum is assumed $^{9,11}$
\begin{equation}\label{lpr1.4}
  P_{A\dot{B}} = \sum\limits^{N}_{i=1} \, \lambda^{(i)}_{A} \,
  \overline{\lambda}^{(i)}_{\dot{B}}\, ,
\end{equation}
where  $P_{A\dot{B}} = \frac{1}{2} \sigma^{\mu}_{AB}P_{\mu}$. We
shall propose analogous formula in $D=11$ for generalized momenta
\begin{equation}\label{lpr1.5}
  Z_{rs} = \sum\limits^{N}_{i=1} \, \lambda^{(i)}_{r} \,
  {\lambda}^{(i)}_{s}\, ,
\end{equation}
where $\lambda_r$ ($r=1 \ldots 32$) are $D=11$ real Majorana
spinors. In $D=4$ the twistors (\ref{lpr1.3}) are the fundamental
representations of the spinorial covering $SU(2,2)$ of $D=4$
conformal algebra ($SU(2,2)=SO\overline{(4,2)}$). In $D=11$ there
exists only minimal conformal spinorial algebra$^{12-14}$
   describing the
 classical real algebra $Sp(64)$, containing $D=11$
conformal algebra
\begin{equation}\label{lpr1.6}
SO\overline{(11,2)} \subset Sp(64;R) \, .
\end{equation}
In Sect. 2 we shall consider the generalization of $D=11$
Poincar\'{e} and conformal superalgebras, supersymmetrizing  the
minimal $D=11$  conformal spinorial algebra. In Sect. 3 we shall
introduce in  $D=11$ the generalization of twistor and
supertwistor formalism, with the extensions of Penrose-Ferber
relations, which relate  $OSp(1|64)$ supertwistor space described
by real  coordinates ($\xi^2 = 0; R=1 \ldots 65$)
\begin{equation}\label{lpr1.7}
  T_R = \left( \lambda_r , \omega^r , \xi \right) \, ,
\end{equation}
with the generalized phase space ($X_{rs}, P_{rs}$) (see
(\ref{lpr1.1}-\ref{lpr1.2})). In Sect. 4 we shall describe
algebraically $ \frac{k}{32} BPS$ states by $n=32-k$ superspinors
(\ref{lpr1.7}) representing $D=11$ generalized supertwistors. These
supertwistorial constituents we shall call BPS preons. It appears
that our  model geometrically corresponds to new type of
Kaluza-Klein theory, with discrete
 internal extension of space-time coordinates.

\section{$D=11$ Conformal $M$-(Super)Algebra}
\vspace*{-0.5pt} \noindent
Let us observe that the $D=4$
conformal algebra ($P_\mu, M_{\mu\nu}, D, K_{\mu}$) is endowed
with the following three - grading structure
\begin{equation}\label{lpr2.1}
  \begin{array}{ccc}
    L_1 & L_0 & L_{-1} \cr\cr
    \label{lpr2.1a}
    P_\mu & M_{\mu\nu}, D & K_{\mu}
    \label{lpr2.1b}
  \end{array} \, .
\end{equation}
Grading (\ref{lpr2.1}) in determined by the scale dimensions of
generators
\begin{equation}\label{lpr2.2}
  [D, P_\mu ] = P_\mu \, ,\qquad
  [D, M_{\mu\nu}] = 0 \, ,
  \qquad
  [ D, K_{\mu}] = - K_{\mu}
\end{equation}
and it is easy to see that the conformal algebra (\ref{lpr2.1}) has two
Poincar\'{e} subalgebras: ($P_{\mu}, M_{\mu\nu}$) and ($K_{\mu},
M_{\mu\nu}$). For $D=4$ superconformal algebra $SU(2,2;1)$ =
$(P_{\mu\nu}$  $M_{\mu\nu}$, $D, A, K_{\mu}$; $Q_{A}$,
$Q_{\dot{A}}$, $S_{A}$,
 $S_{\dot{A}}$) the three-grading (\ref{lpr2.1}) is extended to the
 following five-grading
\begin{equation}\label{lpr2.3}
  \begin{array}{ccccc}
    L_1 & L_{1/2} & L_{0} & L_{- 1/2} &L_{-1}\cr\cr
    P_\mu &Q_{A},\overline{Q}_{\dot{A}}  & M_{\mu\nu}, D,A
     &S_A , \overline{S}_{\dot{A}} & K_{\mu}
  \end{array} \, ,
\end{equation}
where consistently
\begin{equation}\label{lpr2.4}
\begin{array}{ll}
  [D, Q_A ] = \frac{1}{2} Q_A \, ,
\qquad &  [D, S_A ] = - \frac{1}{2} S_A \, ,
   \cr\cr
    [D, \overline{Q}_{\dot{A}} ] = \frac{1}{2} Q_A \, ,
\qquad &     [D, \overline{S}_{\dot{A}} ] =
  \frac{1}{2} \overline{S}_{\dot{A}} \, ,
     \end{array}
\end{equation}
and again $SU(2,2;1)$  contains as subsuperalgebras the
Poincar\'{e} superalgebras  ($P_\mu$, $M_{\mu\nu}$, $Q_{A}$;
$\overline{Q}_{\dot{A}}$) and ($K_\mu, M_{\mu\nu}, S_A,
{S}_{\dot{A}}$).

The structure of $D=11$ generalized superconformal algebra, which
we call conformal $M$-superalgebra is quite analogous. The $D=11$
conformal $M$-algebra $Sp(64)$ can be in analogy to (\ref{lpr2.1})
 described by the following three-grading
\begin{equation}\label{lpr2.5}
  \begin{array}{ccc}
    L_1 & L_0 & L_{-1} \cr\cr
    Z_{rs} & R_{rs} & \widetilde{Z}_{rs}
    \cr\cr
    528 \hbox{\scriptsize \rm \ Abelian }           &GL(32;R)
    & 528 \hbox{\scriptsize \ \rm Abelian }
    \cr
    \hbox{\scriptsize  generators}
    & \hbox{\scriptsize algebra}
    & \hbox{\scriptsize \rm generators}
  \end{array} \, .
\end{equation}
We see that $Sp(64)$ contains two copies of generalized $D=11$
Poincar\'{e} algebras, described by inhomogeneous $Sp(32)$
algebras ($Sp(32;R)\subset GL(32;R)$) with 528 Abelian
translation generators.

The superextension of $D=11$ conformal $M$-algebra $OSp(1;64)$
which we call conformal $M$-superalgebra is described by the
following five-grading (see also$^{15,16}$)
\begin{equation}\label{lpr2.6}
  \begin{array}{ccccc}
    L_1 & L_{1/2} & L_{0} & L_{- 1/2} &L_{-1}\cr\cr
    Z_{rs}  & Q_{r}
    & R_{rs}
     & S_r   & \widetilde{Z}_{rs}
  \end{array} \, ,
\end{equation}
where $(Q_r, S_r)$ are the pair of 32-component supercharges,
transforming as fundamental representations of $Sp(32)$, with
$R_{rs} \subset Sp(32)$ if $R_{rs} = R_{sr}$. The subalgebras
spanned by the generators $(Q_r, Z_{rs})$ and $(S_r,
\widetilde{Z}_{rs})$ describe two copies of $M$-superalgebra
given by the relations (\ref{lpr1.1}).

It should be added that the gradings (12,13) correspond to the
grading structure of real Jordanian (super) algebras [17,18].

\section{$D=11$ Supertwistors and Their Relation with Generalized
Superspace} \vspace*{-0.5pt} \noindent Let us recall two basic
relations of Penrose twistor theory in $D=4$ $^{8-11}$

\begin{romanlist}

\item{} relation between the generators of Poincar\'{e} algebra and
twistor components (2)
\begin{equation}\label{lpr3.1}
  P_{A\dot{B}} = \lambda_A \, \lambda_B\, ,
\end{equation}

\begin{equation}\label{lpr3.2}
  M_{A{B}} =
  \lambda_{(A } \,
   \overline{\omega}_{B )}\, ,
 \qquad  M_{\dot{A} \dot{B}} =
  \overline{\lambda}_{ ( \dot{A}} \,
 \omega_{\dot{B})}
\end{equation}
where $M_{AB} = \frac{1}{2} (\sigma_{\mu\nu})_{AB}M^{\mu\nu}$ and
 $M_{\dot{A}\dot{B}} = \frac{1}{2}
 (\widetilde{\sigma}_{\mu\nu})_{\dot{A}\dot{B}}
 M^{\mu\nu}$\footnote{We recall  that $(\sigma_{\mu\nu})_{AB} = \frac{1}{2i}
 [(\sigma_{\mu})_{A\dot{B}}
  \widetilde{\sigma}_{\nu \ B}^{\ \dot{B}}
 - (\sigma_{\nu})_{A\dot{B}}
  \widetilde{\sigma}_{\mu \  B }^{\dot{B}} ]
 = - \frac{i}{2} \epsilon_{\mu\nu\rho\tau}
  (\sigma^{\rho \tau} )_{AB}
 = [( \widetilde{\sigma}_{\mu\nu})_{\dot{B}\dot{A}}]^{\star}$}.
 The relations (\ref{lpr3.2}) can be extended to all 15 generators
 of $D=4$   conformal algebra.

 \item{} Penrose incidence relation between twistor and space-time
 coordinates
\begin{equation}\label{lpr3.3}
  \omega^{\dot{A}} = i \lambda_{B} X^{B \dot{A}}
  \qquad
   \overline{\omega}^{{A}} = - i  X^{A \dot{B}}\overline{\lambda}_{B}
\end{equation}
where $X^{B \dot{A}} = (X^{A \dot{B}})^{\star}$ describe four
real Minkowski coordinates if the $SU(2,2)$ twistor norm vanishes
\begin{equation}\label{lpr3.4}
  (t, t) \equiv
   i \left(
  \lambda_A \overline{\omega}^{{A}}
  - \overline{\lambda}_{\dot{A}} \omega^{\dot{A}}  \right) = 0\, .
\end{equation}
The relations (\ref{lpr3.1}-\ref{lpr3.4}) can be
supersymmetrized. If we introduce the $D=4$ supertwistor
$(t_{\alpha}, \eta)$, which is the fundamental representation of
$SU(2,2;1)$ with complex Grassmann variable $\eta$ ($\eta^2 =
\overline{\eta}^2 = 0$, $\{ \eta , \overline{\eta} \}= 0$), the
relations (\ref{lpr3.1}-\ref{lpr3.2}) has been extended by
Ferber$^{19}$ to all generators of $D=4$ superconformal group
$SU(2,2;1)$.
\end{romanlist}

The Penrose relations (\ref{lpr3.3}-\ref{lpr3.4}), firstly
supersymmetrized in$^{19}$ look as follows

\begin{eqnarray}\label{lpr3.5}
  \omega^{\dot{A}} & = &
  i \lambda_{B } \, Z^{B\dot{A}}
  \equiv i \lambda_{B} \left(
   X^{B \dot{A}} - i \theta ^{B}
 \theta^{\dot{A}} \right)
 \cr\cr
  \overline{\omega}^A & = &
  i \left( X^{A\dot{B}} + i \theta^A \theta^{\dot{B}} \right)
  \lambda_B
  \cr\cr
  \eta & = & \lambda_{A} \theta^A
   \qquad \overline{\eta} =
 \overline{\lambda}_{\dot{A}} \overline{\theta}^{\dot{A}}\, .
 \end{eqnarray}
 For $D=11$ the generalized twistors and supertwistors are real
 (see \ref{lpr1.7}) and the real $OSp(1;64)$ superalgebra $(R, S = 1 \ldots
 64)$
\begin{equation}\label{lpr3.6}
  \left\{ Q_R , Q_S \right\} = R_{RS}\, ,
\end{equation}
can be obtained if we assume that\footnote{By Bott periodicity
this realization is related with  twistor framework in $D=3$
(see$^{20}$), also with real structure. In $D=5,6,7$ one has to
use the extension of Penrose framework to quaternionic twistors
(see e.g.$^{21}$ for $D=6$).}.
\begin{equation}\label{lpr3.7}
  R_{RS} = T_R \, T_S \qquad
  Q_{R} = \frac{1}{\sqrt{2}} \, T_R \xi\, ,
\end{equation}
where $T_R$ describes $D=11$ real twistorial quantum phase space
($\eta_{RS} = - \eta_{SR}$ is the $Sp(64)$ antisymmetric metric)
\begin{equation}\label{lpt3.8}
  [T_R , T_S ] = i \eta_{RS}\, ,
  \end{equation}
  supplemented   with  trivial one-dimensional Clifford algebra
  relation $\xi^{2} = 1$.

  The relations (\ref{lpr3.6}) are extended to $D=11$ as follows:
\begin{equation}\label{lpr3.9}
  \omega^r = \left( X^{rs} - i \theta^r \, \theta^s \right)
  \lambda_s \qquad \xi= \theta^r \, \lambda_r \, .
\end{equation}
Relations (\ref{lpr3.9}) relate the $D=11$ supertwistor space
coordinates (\ref{lpr1.7}) with the extended $D=11$ superspace
$(X_{rs}, \theta_s)$, described by 528 bosonic and 32 fermionic
coordinates.

\section{BPS States in $M$-Theory and Composites of BPS Preons}
\vspace*{-0.5pt} \noindent The $\frac{k}{32}BPS$ state $
|k\rangle$ can be defined as an eigenstate of generalized momenta
generators
\begin{equation}\label{lpr4.1}
  Z_{rs} | k\rangle = z_{rs} | k\rangle \, ,
\end{equation}
such that $\det z_{rs} = 0$. The number $k$ determines the rank of
generalized momenta matrix $z_{rs}$
\begin{equation}\label{lpr4.2}
  \frac{k}{32} \hbox{BPS\  state:} \left\{
  \hbox{rank}\ z_{rs} = n = 32 - k; \quad 1\leq k < 32 \right\}\, .
\end{equation}
From (\ref{lpr4.2}) follows that the $BPS$ state $|k\rangle$
preserves a fraction $\nu = \frac{k}{32}$ of supersymmetries.

We call $BPS$ preon the hypothetical primary object carrying the
following generalized momenta$^{1}$
\begin{equation}\label{lpr4.3}
  Z_{rs} = \lambda_r \, \lambda_s\, .
\end{equation}
The formula (\ref{lpr4.3}) corresponds to putting $n=1$ in the
relation (\ref{lpr1.5}) and describes $\frac{31}{32}BPS$ state.
More general
 formula (\ref{lpr1.5}) describes the generalized momenta of a system
composed out of $n$ $BPS$ preons and it describes (for $1 \leq n
\leq 32$) the $\frac{k}{32} BPS$ state (we recall that $k=32 -n
$).

The number $n=32-k$ of zero eigenvalues of the matrix $z_{rs}$
determines the number of independent supercharges $Q^{(i)}_r$,
anihilating the $BPS$ state $|k\rangle $. These supersymmetries,
preserving the $BPS$ state, are called in $p$-brane theory the
$\kappa$-transformations. We see that the supersymmetric $D=11$
single $BPS$ preon dynamics should have 31 $\kappa$-symmetries.
Recently$^{22}$ such dynamical superparticle models\footnote{For
$D=4,6$ and 10 see$^{22,23}$.} with fundamental $OSp(1;2n)$
superspinor as basic variable has been proposed. It should be
recalled here (see e.g.$^{24}$) that in the standard super
$p$-brane formulations half of the supersymmetries are promoted to
$\kappa$-transformations, i.e. in $D=11$ we obtain 16
$\kappa$-transformations.

Using the $D=11$ supertwistor description with the relations
(\ref{lpr3.9}) and (\ref{lpr4.3}) providing a bridge between
$BPS$ preons and  generalized space-time, we can formulate three
different geometric pictures:

\begin{romanlist}
\item{} Purely supertwistorial picture, with basic phase
space parametrized by $BPS$ preon coordinates $T^{(i)}_R$ (see
(\ref{lpr1.7})). The canonical Liouville one-form describing free
action is given by the relation
\begin{equation}\label{lpr4.4}
  \Omega_1 =
  \sum\limits^{n}_{i=1}
\left(
   \omega^{(i)r} \,
  d\lambda^{(i)}_{\ r}
  + i \xi^{(i)} \, d \xi^{(i)}\right) \, ,
\end{equation}
which can be supplemented by some algebraic constraints.

\item{} Mixed geometric picture, with the components
$\omega^{(i)}$ expressed by means of  the relation (\ref{lpr3.9}).
One obtains from (\ref{lpr4.4})
\begin{equation}\label{lpr4.5}
  \Omega_2 =
   \sum\limits^{n}_{i=1} \lambda^{(i)r}_{\ r} \,
  \lambda^{(i)r}_{\ s}
\left(  dX^{rs} - i \theta^{r} \, d\theta^{s}\right) \, ,
\end{equation}

\item{} Generalized space-time picture, with the relation
(\ref{lpr1.5}) inserted in (\ref{lpr4.5}).
\begin{equation}\label{lpr4.6}
  \Omega_3 =
  Z_{rs}
\left(  dX^{rs} - i \theta^{r} \, d\theta^{s}\right) \, .
\end{equation}
The application of these three geometric pictures to the
description of $D=11$ dynamics (for $n>1$) is under consideration.
\end{romanlist}

\section{Final Remarks}
\vspace*{-0.5pt} \noindent We mention here two interesting
aspects of the presented approach which deserve further attention;

\begin{romanlist}
\item{geometric confinement of $BPS$ preons}

Because the space-time coordinates are composed out of preonic
degrees of freedom, the $D=11$ space-time point can be determined
only in terms of at least 16 preonic set of spinorial coordinates.
This is the $D=11$ extension of known property of Penrose theory
in four dimensions with two twistors needed for the definition of
composite Minkowski space-time points.

\item{internal symmetries}

The formula (\ref{lpr1.5}) expresses 528 generalized momenta in
terms of 32$n$ preonic spinorial coordinates $\lambda^{(i)}_{\ r}$
 ($i=1, \ldots n$).
 The internal symmetries can be obtained by interchanging $BPS$ preons.
  For the case n=16 corresponding to
  the choice of  $\nu = \frac{1}{2}
 \hbox{SUSY}$ one can introduce internal $O(16)$ symmetries,
 leaving the values of $Z_{rs}$ invariant.
\end{romanlist}

\nonumsection{Acknowledgements} \noindent The author would like
thank prof. Mo-Lin Ge for his warm hospitality at Nankai Symposium
in Tianjin.

\nonumsection{References}

\end{document}